\documentclass[prd,preprintnumbers,superscriptaddress,twocolumn,epsf,nofootinbib,showpacs]{revtex4-2}

\usepackage{amsfonts,amssymb,amsmath}
\usepackage{mathrsfs}
\usepackage{color}
\usepackage{graphicx}
\usepackage{dcolumn}
\usepackage{multirow}
\usepackage{bm}
\usepackage{pslatex}
\usepackage[T1]{fontenc}
\usepackage[latin1]{inputenc}

\begin{document}

%%%%%%%%%%%%%%%%%%%%%%%%
%%%%%%%%%%%%%%%%%%%%%%%%
\title{General Analytical Conditions for Inflaton Fragmentation: Quick and Easy Tests for its Occurrence}

\author{Jinsu Kim}
\email{jinsu.kim@cern.ch}
\affiliation{
	Theoretical Physics Department,
	CERN,
	1211 Geneva 23, Switzerland
}

\author{John McDonald}
\email{j.mcdonald@lancaster.ac.uk}
\affiliation{
	Department of Physics, 
	Lancaster University, 
	Lancaster LA1 4YB, United Kingdom
}

\date{\today}

\preprint{CERN-TH-2021-202}

\begin{abstract}
Understanding the physics of inflaton condensate fragmentation in the early Universe is crucial as the existence of fragments in the form of non-topological solitons (oscillons or Q-balls) may potentially modify the evolution of the post-inflation Universe. Furthermore, such fragments may evolve into primordial black holes and form dark matter, or emit gravitational waves. Due to the non-perturbative and non-linear nature of the dynamics, most of the studies rely on numerical lattice simulations. Numerical simulations of condensate fragmentation are, however, challenging and, without knowing where to look in the parameter space, they are likely to be time-consuming as well. In this paper, we provide generic analytical conditions for the perturbations of an inflaton condensate to undergo growth to non-linearity in the cases of both symmetric and asymmetric inflaton potentials. We apply the conditions to various inflation models and demonstrate that our results are in good agreement with explicit numerical simulations. Our analytical conditions are easy to use and may be utilised in order to quickly identify models that may undergo fragmentation and determine the conditions under which they do so, which can guide subsequent in-depth numerical analyses.
\end{abstract}

%\pacs{xxx}
%\keywords{xxx}
\maketitle
%%%%%%%%%%%%%%%%%%%%%%%%
%%%%%%%%%%%%%%%%%%%%%%%%

%%%%%%%%%%%%%%%%%%%%%%%%
\section{Introduction}
\label{sec:intro}
%%%%%%%%%%%%%%%%%%%%%%%%

After inflation ends, a real scalar inflaton field starts to oscillate around the minimum of its potential. An attractive inflaton self-interaction may then result in growth of perturbations of the inflaton to non-linearity and, if non-topological soliton (NTS) solutions exist, which will be the case in the models we consider, subsequent fragmentation of the inflaton condensate\footnote{
It is important to distinguish between the growth of perturbations of a scalar field to non-linearity and condensate fragmentation. The latter requires that discrete NTS solutions exist, so that the condensate may break up into discrete NTS with vacuum between them. The condition for NTS to be possible in models that grow to non-linearity is that the scalars are sufficiently massive in the vacuum so that the energy per scalar in the NTS can be less than the mass of the scalar in the vacuum. This will be generally true of the potentials that we will consider.
}. The inflaton condensate fragments correspond to oscillons, which are spherically symmetric quasi-stable NTS \cite{McDonald:2001iv}. The longevity of oscillons can be understood by the conservation of adiabatic charge \cite{Kasuya:2002zs}.

The study of inflaton condensate fragmentation has gained much attention partly due to the interesting consequences of the fragments. Examples include primordial black hole formation due to the statistics of fragments \cite{Cotner:2018vug,Cotner:2019ykd,Flores:2021tmc}, gravitational wave signals associated with fragments \cite{Zhou:2013tsa,Antusch:2017vga,Liu:2018rrt,Sang:2019ndv,White:2021hwi}, and changes in reheating dynamics which lead to a different cosmic evolution \cite{Lozanov:2016hid,Lozanov:2017hjm}.

The evolution dynamics of the fragments often requires numerical lattice simulations due to the non-perturbative and non-linear nature. Most of the existing literature adopts numerical approaches and performs extensive lattice simulations. Numerical simulations are, however, often technically challenging and computationally demanding, and can obscure the underlying physics. Furthermore, without knowing where to look in the model parameter space, or if fragmentation is even possible in a given model, numerical studies can be time-consuming. Having an analytical expression for the condition under which the condensate is likely to undergo non-linear growth and hence, if discrete NTS solutions exist, fragmentation, would be greatly beneficial for such in-depth numerical analyses. Such a condition could serve as a starting point of extensive numerical analyses, and it could also be used to quickly estimate the likelihood of fragmentation in a given model. We expect that NTS will exist in the models that we consider, therefore, in the following, we will consider the growth of perturbations to non-linearity and fragmentation to be equivalent.

The aim of this paper is to provide general analytical conditions for perturbations of an inflaton condensate to grow to become non-linear and hence for inflaton condensate fragmentation to be possible.
In Ref.~\cite{Kim:2017duj} we studied the stability of an inflaton condensate, based on Ref.~\cite{McDonald:2001iv}, for the case of a general symmetric potential. We obtained analytical conditions on the self-interaction couplings under which the condensate undergoes fragmentation. However, the method of Ref.~\cite{Kim:2017duj} cannot be used when the potential is asymmetric. In this work, we generalise the analysis of Ref.~\cite{Kim:2017duj} to encompass asymmetric potentials, thus allowing the condition for fragmentation to be determined for a wide range of inflaton potentials.

In the next section, we develop an analytical framework for studying the growth of inflaton field perturbations and inflaton condensate fragmentation. We then provide the analytical conditions under which the condensate perturbations grow to non-linearity and fragment for both symmetric and asymmetric potentials. The main results of the paper are summarised in Sec. \ref{subsec:ACsummary}. In Sec. \ref{sec:apps}, we apply our analytical results to four examples for which numerical results are available: the $\alpha$-attractor $T$-model, the $\alpha$-attractor $E$-model, Starobinsky's $R^2$ model, and the Palatini $R^2$ model with a quadratic potential. In addition, we apply our results to a model which has not yet been studied numerically, Higgs Inflation with a general symmetry-breaking potential in both the metric and Palatini formulations, and we perform lattice simulations to test our analytical results. We show that our results are in good agreement with numerical simulations in all cases, thus demonstrating their general effectiveness. We conclude in Sec. \ref{sec:conc}.

%%%%%%%%%%%%%%%%%%%%%%%%
\section{General Treatment}
\label{sec:GenTreat}
%%%%%%%%%%%%%%%%%%%%%%%%

%%%%%%%%%%%%%%%%%%%%%%%%
\subsection{Inflaton condensate}
\label{subsec:InfConden}
%%%%%%%%%%%%%%%%%%%%%%%%
Consider a real scalar field $\Phi$ whose equation of motion\footnote{
Throughout the paper, we work in the Einstein frame with a canonically normalised kinetic term, unless otherwise stated.
} is given by
\begin{align}
\ddot{\Phi} + 3H\dot{\Phi} - \frac{\nabla^{2}\Phi}{a^{2}} + V_{\Phi} = 0
\,,
\end{align}
where the dot denotes the cosmic time derivative and $V_{\Phi} \equiv dV(\Phi)/d\Phi$.

In our analysis, we will be considering a potential that is close to a quadratic potential, so to a first approximation the field will undergo coherent oscillations in a quadratic potential. We expect that any potential that is capable of supporting NTS solutions will be approximately quadratic at the minimum of its potential, as it is necessary for the scalars to have a non-zero mass in the vacuum in order for the NTS solutions to have a lower energy per scalar and so to be (meta)stable.

Defining $\Phi = (a_{0}/a)^{3/2}\phi$, where we have assumed that the coherent oscillations start at $a=a_{0}$, gives
\begin{align}\label{eqn:eomphi}
\ddot{\phi} - \frac{\nabla^{2}\phi}{a^{2}}
=
-\frac{\partial U(\phi)}{\partial \phi}
\,,
\end{align}
where
\begin{align}
\frac{\partial U}{\partial \phi} \equiv
\left(\frac{a}{a_{0}}\right)^{3/2}V_{\Phi} + \Delta_{H}\phi
\,,
\end{align}
with
\begin{align}\label{eqn:DeltaHexpression}
\Delta_{H} \equiv -\frac{3}{2}\left(\dot{H} + \frac{3}{2}H^{2}\right)\,.
\end{align}

In the case of a symmetric potential, $V(\Phi) = V(-\Phi)$, the oscillatory behaviour of the scalar field can well be described by $\phi(t,\mathbf{x}) = R \cos\Omega$, where $R$ and $\Omega$ are both functions of $t$ and $\mathbf{x}$. The inflaton condensate fragmentation in symmetric potentials is studied in Ref.~\cite{Kim:2017duj}.
However, for asymmetric potentials, $V(\Phi) \neq V(-\Phi)$, the analysis of Ref.~\cite{Kim:2017duj} cannot be applied as the choice of $\phi = R \cos\Omega$ does not capture the asymmetric feature.
In the case of an asymmetric potential, we expect the amplitude of the oscillating field to be different at $\Omega = 0$ and $\Omega = \pi$.
Assuming that the potential is dominated by a quadratic term, one may model this by considering
\begin{align}\label{eqn:phiFormGen}
\phi = R(1+\epsilon \cos\Omega)\cos\Omega\,,
\end{align}
where $\epsilon \ll 1$ parametrises the asymmetry of the potential. The form of $\epsilon$ can be determined once the potential is specified. We explore the case of an asymmetric potential in more detail in Sec. \ref{subsec:asymmpot}. Note that the symmetric potential case corresponds to the $\epsilon\rightarrow 0$ limit. Thus, Eq. \eqref{eqn:phiFormGen} may be used to describe the inflaton condensate motion for a generic form of potential.

The equation of motion \eqref{eqn:eomphi} can be expressed in terms of $R=R(t,\mathbf{x})$ and $\Omega=\Omega(t,\mathbf{x})$ as
\begin{align}\label{eqn:eomROmega}
0 &=
\ddot{R}(1+\epsilon\cos\Omega)\cos\Omega
-2\dot{R}\dot{\Omega}(1+2\epsilon\cos\Omega)\sin\Omega
\nonumber\\
&\quad
-R\ddot{\Omega}(1+2\epsilon\cos\Omega)\sin\Omega
-R\dot{\Omega}^2(\cos\Omega + 2\epsilon\cos(2\Omega))
\nonumber\\
&\quad
-\frac{\nabla^2R}{a^2}(1+\epsilon\cos\Omega)\cos\Omega
\nonumber\\
&\quad
+\frac{2(\nabla R \cdot \nabla\Omega)}{a^2}(1+2\epsilon\cos\Omega)\sin\Omega
\nonumber\\
&\quad
+\frac{R(\nabla^2\Omega)}{a^2}(1+2\epsilon\cos\Omega)\sin\Omega
\nonumber\\ &\quad
+\frac{R(\nabla\Omega)^2}{a^2}(\cos\Omega+2\epsilon\cos(2\Omega))
+\frac{\partial U}{\partial\phi}
\,.
\end{align}
Multiplying the equation of motion \eqref{eqn:eomROmega} by $\sin\Omega$ and averaging over coherent oscillations, we obtain
\begin{align}\label{eqn:eom1}
\ddot{\Omega}
+2\dot{R}\dot{\Omega}/R
-2\nabla R \cdot \nabla\Omega / (a^{2}R)
-(\nabla^{2}\Omega)/a^{2} = 0\,.
\end{align}
Similarly, multiplying the equation of motion by $\cos\Omega$ and averaging over oscillations, we obtain
\begin{align}\label{eqn:eom2}
&\ddot{R} - R\dot{\Omega}^{2}
-\nabla^{2}R/a^{2}
+R(\nabla\Omega)^{2}/a^{2}
+U_{\rm eff}^\prime
=0
\,,
\end{align}
where $U_{\rm eff}^\prime \equiv 2\langle \cos\Omega \frac{\partial U}{\partial\phi} \rangle$.
Here, we have assumed that $R$ and $\dot{\Omega}$ do not change much over the period of oscillations, which is a good approximation when the potential is dominated by the quadratic term.
Furthermore, we can assume that $\Delta_{H}$ term is small enough to be ignored, which is a good approximation if $\omega$ is large compared to $H$, where $\omega$ is the oscillation frequency, because in Eq. \eqref{eqn:eomphi} $\Delta_H \sim H^2$ is effectively a contribution to the frequency squared of the oscillations and so is negligible if $\omega^2 \gg H^2$.
We note that Eqs. \eqref{eqn:eom1} and \eqref{eqn:eom2} have the same form as those studied in Ref. \cite{Kim:2017duj}. We can thus follow the same procedure as Ref. \cite{Kim:2017duj}. For completeness, we repeat the procedure.

We expand $R$ and $\Omega$ into their background parts and perturbations parts as
\begin{align}\label{eqn:ROmPertParts}
R(t,\mathbf{x}) = R_0(t) + \delta R(t,\mathbf{x}) \,,\;
\Omega(t,\mathbf{x}) = \Omega_0(t) + \delta \Omega(t,\mathbf{x})\,.
\end{align}
with $R_{0}$ and $\Omega_{0}$ being the solutions of Eqs. \eqref{eqn:eom1} and \eqref{eqn:eom2}.
It is straightforward to show that the perturbations satisfy
\begin{align}
\ddot{\delta R} - 2R_0\dot{\Omega}_0\dot{\delta\Omega}
-\dot{\Omega}_0^2\delta R - \frac{\nabla^2\delta R}{a^2} 
+U_{\rm eff}^{\prime\prime}\delta R  &= 0
\,,\label{eqn:perteom1}\\
\ddot{\delta\Omega} - \frac{\nabla^2\delta\Omega}{a^2}
-\frac{2\dot{R}_0\dot{\Omega}_0}{R_0^2}\delta R
+\frac{2\dot{R}_0}{R_0}\dot{\delta \Omega}
+\frac{2\dot{\Omega}_0}{R_0}\dot{\delta R} &= 0 \,,\label{eqn:perteom2}
\end{align}
where we assumed that the background solutions $R_{0}$ and $\Omega_{0}$ are homogeneous and we considered terms only up to the first order in $\delta R$ and $\delta \Omega$.
The solution for the growth modes takes the form \cite{McDonald:2001iv}
\begin{align}
\delta R, \delta \Omega
\propto e^{S(t) - i\mathbf{k}\cdot \mathbf{x}}
\,,
\end{align}
where we may assume that $s \equiv \dot{S} = \text{constant}$ on timescales short compared to the expansion time since $|\dot{s}/s| \sim H$ \cite{McDonald:2001iv}.
Thus, assuming the oscillation time scale is short compared to the expansion time $H^{-1}$, we obtain
\begin{align}
0=\left[
s^{2}
+\frac{k^{2}}{a^{2}}
+U_{{\rm eff}}^{\prime\prime} - \dot{\Omega}_{0}^{2}
\right]\left[
s^{2}
+\frac{k^{2}}{a^{2}}
\right]
+4s^{2}\dot{\Omega}_{0}^{2}
\,.
\end{align}
One can solve the equation for $s^{2}$,
\begin{align}
s^{2} &=
-\frac{k^{2}}{a^{2}}
-\frac{3\dot{\Omega}_{0}^{2}+U_{{\rm eff}}^{\prime\prime}}{2}
\nonumber\\
&\quad
\pm\left(
\frac{|3\dot{\Omega}_{0}^{2}+U_{{\rm eff}}^{\prime\prime}|}{2}
\right)\left(
1
+\frac{16 (k^{2}/a^{2}) \dot{\Omega}_{0}^{2}}{(3\dot{\Omega}_{0}^{2}+U_{{\rm eff}}^{\prime\prime})^{2}}
\right)^{1/2}
\,.
\end{align}
For small $k/a$ values,
\begin{align}
s^{2} &\approx
-\frac{k^{2}}{a^{2}}
-\frac{3\dot{\Omega}_{0}^{2}+U_{{\rm eff}}^{\prime\prime}}{2}
\\&
\pm\left(
\frac{|3\dot{\Omega}_{0}^{2}+U_{{\rm eff}}^{\prime\prime}|}{2}
\right)\left(
1 + \frac{8 (k^{2}/a^{2}) \dot{\Omega}_{0}^{2}}{(3\dot{\Omega}_{0}^{2}+U_{{\rm eff}}^{\prime\prime})^{2}}
-\frac{36 (k^{4}/a^{4}) \dot{\Omega}_{0}^{4}}{(3\dot{\Omega}_{0}^{2}+U_{{\rm eff}}^{\prime\prime})^{4}}
\right)\nonumber
\,.
\end{align}
For a growing mode, $s^{2} > 0$, and thus we take the $+$ sign, provided that $3\dot{\Omega}_{0}^{2}+U_{{\rm eff}}^{\prime\prime} > 0$, which gives
\begin{align}\label{eqn:alphasoln}
s^{2} &\approx
\frac{k^{2}}{a^{2}}
\frac{1}{3\dot{\Omega}_{0}^{2}+U_{{\rm eff}}^{\prime\prime}}
\left[
\dot{\Omega}_{0}^{2}-U_{{\rm eff}}^{\prime\prime}
-16\frac{k^{2}}{a^{2}}\frac{\dot{\Omega}_{0}^{4}}{(3\dot{\Omega}_{0}^{2}+U_{{\rm eff}}^{\prime\prime})^{2}}
\right]
\,.
\end{align}
The largest possible value for $k^{2}/a^{2}$ is then given by
\begin{align}\label{eqn:kmax}
\frac{k_{{\rm max}}^{2}}{a^{2}} \approx
\dot{\Omega}_{0}^{2}-U_{{\rm eff}}^{\prime\prime}
\,.
\end{align}
Note that $\dot{\Omega}_{0}^{2} = U_{{\rm eff}}^{\prime}/R$ from the background equation of motion.
We therefore obtain
\begin{align}
s^{2} \approx \frac{k^{2}}{a^{2}}\left(
1 - \frac{k^{2}}{k_{{\rm max}}^{2}}\right)
\frac{U_{{\rm eff}}^{\prime}/R - U_{{\rm eff}}^{\prime\prime}}{3U_{{\rm eff}}^{\prime}/R+U_{{\rm eff}}^{\prime\prime}}
\,.
\end{align}
We now can integrate $s$ to obtain the total growth factor,
\begin{align}\label{eqn:growthfactor}
S(k,a(t)) =
\int_{t_{0}}^{t} dt \, s(k,t)
=
\int_{a_{0}}^{a(t)} da \,
\frac{s(k,a)}{aH}
\,.
\end{align}

As a condition for fragmentation, we require $S \sim \ln(R_0/\delta R_0)$, which results in non-linear perturbations, i.e., $\delta R/R \sim 1$. The primordial inflaton perturbation, $\delta R_0/R_0 \sim 10^{-4}$, can be used as the initial perturbation. Thus, the condition for growth to non-linearity and so fragmentation is given by
\begin{align}\label{eqn:condFrag}
S(k,a(t)) > 10\,.
\end{align}

The discussion above is applicable for a general potential. The polynomial expansion of a general potential is expected to have a leading-order quartic or cubic interactions for a symmetric potential or an asymmetric potential, respectively. The fragmentation condition \eqref{eqn:condFrag} can then be converted into a bound on the model parameters. In the following two subsections, we discuss the case of a symmetric and an asymmetric potential respectively and obtain the condition on the model parameters for fragmentation to occur.

%%%%%%%%%%%%%%%%%%%%%%%%
\subsection{Symmetric potentials}
\label{subsec:symmpot}
%%%%%%%%%%%%%%%%%%%%%%%%
Let us consider
\begin{align}\label{eqn:PotSymm}
V(\Phi) = \frac{1}{2}m^2\Phi^2 - A\Phi^4\,,
\end{align}
where $A>0$. The quadratic term is assumed to be dominant. The leading-order quartic self-interaction is expected to naturally arise in the polynomial expansion of a general symmetric potential\footnote{
For the potential of the form $V(\Phi) = \frac{1}{2}m^2\Phi^2 - A|\Phi|^3$, one may refer to Ref. \cite{Kim:2017duj}.
}. The condition for fragmentation to occur with the quartic self-interaction was studied in detail in Ref. \cite{Kim:2017duj}. For completeness, we repeat the calculation here.

As the potential is symmetric under $\Phi \rightarrow -\Phi$, we may set $\epsilon = 0$ in Eq. \eqref{eqn:phiFormGen} and obtain the following effective potential,
\begin{align}
U_{\rm eff}(R) = \frac{1}{2}m^2R^2 - \frac{3}{4}A\left(
\frac{a_0}{a}
\right)^3 R^4\,.
\end{align}
The upper limit of the comoving wavenumber, $k_{\rm max}$, is then given by Eq. \eqref{eqn:kmax},
\begin{align}
k_{\rm max} \approx \sqrt{6A} R a_0^{3/2} a^{-1/2} \,.
\end{align}
We note that $k_{\rm max} \propto a^{-1/2}$. For a mode $k<k_{\rm max}$, the mode is initially growing. Eventually, the mode $k$ will become larger than $k_{\rm max}$ since $k_{\rm max}$ decreases as $a$ increases. Therefore, the mode will stop growing.

The maximum possible growth of a mode can be found by first considering perturbations with $k<k_{\rm max,0}$ at an initial scale factor $a_0$. 
We then calculate the growth of the perturbations from $a_0$ to a later scale factor $a_{\rm f}$, $S(k,a_{\rm f})$, noting that the perturbation growth for a given $k$ will stop once $k>k_{\rm max}(a_{\rm f})$.
We maximise $S(k,a_{\rm f})$ with respect to $k$ for each $a_{\rm f}$, corresponding to $k(a_{\rm f})$, and finally we maximise $S(k(a_{\rm f}), a_{\rm f})$ with respect to $a_{\rm f}$.
In this way we determine the maximum possible growth of a mode starting at $a_0$.

Using Eq. \eqref{eqn:growthfactor}, we obtain the growth factor from $a_0$ to $a_{\rm f}$ as follows:
\begin{align}
S(k,a_{\rm f}) = \sqrt{\frac{3AR^2}{2m^2}}
\frac{k}{a_0 H_0}
\int_1^{x_{\rm f}} x^{-2} \left(
1 - \frac{k^2}{k_{\rm max,0}^2}x
\right)^{1/2} dx\,,
\end{align}
where $x_{\rm f} \equiv a_{\rm f}/a_0$.
A conservative analytical result for the growth factor can be obtained by setting $x = x_{\rm f}$ in the square root factor, as it overestimates the suppression of the growth factor. Then we find, with $k_{\rm max,f} \equiv k_{\rm max}(a_{\rm f})$,
\begin{align}
S(k,a_{\rm f}) =
\sqrt{\frac{3AR^2}{2m^2}}
\frac{k}{a_0 H_0}
\left(
1 - \frac{k^2}{k_{\rm max,f}^2}
\right)^{1/2}
\left(
1 - \frac{a_0}{a_{\rm f}}
\right)\,.
\end{align}
Maximising the growth factor with respect to $k$ gives a maximum at $k(a_{\rm f})$, where
\begin{align}
k(a_{\rm f}) = \frac{k_{\rm max,f}}{\sqrt{2}} = \left(\frac{a_0}{a_{\rm f}}\right)^{1/2}
\frac{k_{\rm max,0}}{\sqrt{2}}\,.
\end{align}
Thus,
\begin{align}
S(k(a_{\rm f}),a_{\rm f}) = \sqrt{\frac{3AR^2}{2m^2}}
\frac{k_{\rm max, 0}}{2a_0 H_0}
\left(\frac{a_0}{a_{\rm f}}\right)^{1/2}
\left(
1 - \frac{a_0}{a_{\rm f}}
\right)\,.
\end{align}
Maximising this with respect to $a_{\rm f}$ gives
\begin{align}
a_{\rm f} = 3 a_0 \,.
\end{align}
Therefore,
\begin{align}
S_{\rm max} = \frac{1}{3\sqrt{3}}
\sqrt{\frac{3AR^2}{2m^2}}
\frac{k_{\rm max, 0}}{a_0 H_0}
= \frac{\sqrt{2}ARM_{\rm P}}{m^2}\,,
\end{align}
where we used the fact that the potential is dominated by the quadratic term so that $H_0 = mR/(\sqrt{6}M_{\rm P})$. We should choose $R \equiv \Phi_0$, where $\Phi_0$ is the amplitude of the field oscillations at $a_0$, to be as large as possible while being consistent with the assumption that the potential is dominated by the quadratic term. We define the ratio between the quartic part and the quadratic part of the potential by $r_S$,
\begin{align}\label{eqn:rS}
r_S = \frac{A\Phi_0^4}{m^2\Phi_0^2/2} = \frac{2AR^2}{m^2}\,.
\end{align}
In terms of $r_S$, we then obtain
\begin{align}
S_{\rm max} = (r_S A)^{1/2}\frac{M_{\rm P}}{m}\,.
\end{align}
Therefore, the fragmentation condition \eqref{eqn:condFrag}, $S_{\rm max}>10$, will be satisfied if
\begin{align}
A > 1000\left(\frac{0.1}{r_S}\right)\left(\frac{m}{M_{\rm P}}\right)^2\,.
\end{align}
We consider $r_S \approx 0.1$ in order to be consistent with our assumption that the quadratic potential term is dominant while still making $r_S$ sufficiently large to give the smallest possible lower bound on the coefficient of the quartic potential term for fragmentation to occur \footnote{This also assumes that $r_S > 0.1$ at the end of inflation, which is typically true.}.

%%%%%%%%%%%%%%%%%%%%%%%%
\subsection{Asymmetric potentials}
\label{subsec:asymmpot}
%%%%%%%%%%%%%%%%%%%%%%%%
Next, we consider a typical asymmetric potential which, in the small field region where inflaton starts to oscillate around the minimum of the potential after inflation ends, can generically be described by the leading-order cubic term,
\begin{align}\label{eqn:PotAsymm}
V(\Phi) = \frac{1}{2}m^{2}\Phi^{2} \pm A\Phi^{3} \,.
\end{align}
The quadratic term is assumed to be the dominant term in the potential as before. The sign of the cubic term is chosen in such a way that $A$ takes a positive value.

In the case of asymmetric potentials, $\epsilon$ in Eq. \eqref{eqn:phiFormGen} takes a non-zero value. To find $\epsilon$, let us assume that the field reaches maximum at $\Phi=\Phi_{\pm}$ with $\dot{\Phi} = 0$, where $\Phi_{\pm} = \pm(\Phi_{0}\pm \delta\Phi)$ for the $-$ sign in the potential \eqref{eqn:PotAsymm} and $\Phi_{\pm} = \pm(\Phi_{0}\mp \delta\Phi)$ for the $+$ sign in the potential \eqref{eqn:PotAsymm}. Here, $\Phi_{0}, \delta\Phi > 0$. Note that $\delta\Phi \rightarrow 0$ as $A\rightarrow 0$.
From $V(\Phi_{+}) = V(\Phi_{-})$, we find that $\delta\Phi = A\Phi_{0}^{2}/m^{2}=AR^{2}/m^{2}$, where $\Phi_{0}$ is identified with $R$.
Thus, the parameter $\epsilon$ can be determined as
\begin{align}\label{eqn:epsilonDef}
\epsilon = \pm\frac{AR}{m^{2}}\,,
\end{align}
with the $+$ ($-$) sign for the $-$ ($+$) sign in the potential \eqref{eqn:PotAsymm}, representing the deviation from the quadratic term.

The effective potential now takes
\begin{align}\label{eqn:Ueff}
U_{{\rm eff}} =
\frac{1}{2}m^{2}R^{2} - \frac{9A^{2}}{8m^{2}}
\left(
\frac{a_{0}}{a}
\right)^{3/2}R^{4}
\,.
\end{align}
The upper limit of the comoving wavenumber, $k_{\rm max}$, in this case is given by
\begin{align}
k_{\rm max} \approx \frac{3AR}{m}a_0^{3/4}a^{1/4}\,.
\end{align}
We see that $k_{\rm max} \propto a^{1/4}$. Thus, a wavenumber that is initially larger than the maximum wavenumber, $k>k_{\rm max}$, will eventually become smaller than $k_{\rm max}$, and the growth of the mode starts. Therefore, the maximum growth corresponds to a mode $k=k_{\rm max,0}$, for a given $a_0$.
The growth factor is given by
\begin{align}
S(k_{\rm max,0},a) = \frac{3AR}{2m^{2}}\frac{k_{{\rm max,0}}}{a_0 H_0}
\int_{1}^{x} 
x^{-5/4}\sqrt{1-x^{-1/2}}  \, dx
\,,
\end{align}
where $x \equiv a/a_0$.
We take the limit $x \rightarrow \infty$. Then, the maximum growth factor is given by
\begin{align}
S_{{\rm max}} = \frac{9\sqrt{6}\pi A^{2}RM_{{\rm P}}}{2m^{4}}
\,,
\end{align}
where we used the fact that the potential is dominated by the quadratic term so that $H_0 = mR/(\sqrt{6}M_{{\rm P}})$.

Therefore, the fragmentation condition \eqref{eqn:condFrag}, $S_{\rm max}>10$, will be satisfied if
\begin{align}\label{eqn:fragCond1}
\frac{A^{2} R M_{{\rm P}}}{m^{4}} 
>
\frac{20}{9\sqrt{6}\pi}
\approx
0.29\,.
\end{align}
We may take the value of $R$ as large as possible while keeping the cubic term in the potential smaller than the quadratic term. We parametrise it as $r_A$, similarly to the $r_S$ in the symmetric case \eqref{eqn:rS},
\begin{align}
R = r_A \frac{m^{2}}{2A}\,.
\end{align}
Thus, the condition for growth to non-linearity and so fragmentation \eqref{eqn:fragCond1} can be written as
\begin{align}\label{eqn:fragCond2}
A > \frac{400}{9\sqrt{6}\pi}\left(
\frac{m^{2}}{M_{{\rm P}}}
\right)\left(
\frac{0.1}{r_A}
\right)
\approx 5.78 \left(
\frac{m^{2}}{M_{{\rm P}}}
\right)\left(
\frac{0.1}{r_A}
\right)\,.
\end{align}
Similarly to the symmetric potential case, we take $r_A \approx 0.1$ in order to be consistent with our assumption that the quadratic potential term is dominant while giving the smallest possible lower bound on the coefficient of the cubic potential term for fragmentation to occur.

%%%%%%%%%%%%%%%%%%%%%%%%
\subsection{Summary of analytical conditions}
\label{subsec:ACsummary}
%%%%%%%%%%%%%%%%%%%%%%%%
We summarise the general analytical fragmentation conditions:
\begin{itemize}
\item Symmetric potentials $V = \frac{1}{2}m^2\Phi^2 - A\Phi^4$:
\begin{align}\label{eqn:MainResultSymm}
A > 1000\left(\frac{m}{M_{\rm P}}\right)^2\left(\frac{0.1}{r_S}\right)\,.
\end{align}

\item Asymmetric potentials $V = \frac{1}{2}m^2\Phi^2 \pm A\Phi^3$:
\begin{align}\label{eqn:MainResultAsymm}
A >\frac{400}{9\sqrt{6}\pi} \left(\frac{m^{2}}{M_{{\rm P}}}\right)\left(\frac{0.1}{r_A}\right)\,.
\end{align}
\end{itemize}
These are the main results of the paper.

Strictly speaking, these conditions are sufficient conditions for fragmentation, as they show that if the condensate forms it will fragments, hence fragmentation is guaranteed if our conditions are satisfied. In our analysis, we are restricting the initial value of the field to be close enough to the minimum of the potential that the potential is dominated by the quadratic term. However, it is possible that fragmentation could occur more rapidly and at larger field values due to tachyonic preheating, in which case it is possible that fragmentation could still occur even if our conditions are not satisfied. In practice, as we will demonstrate via a range of examples below, we find that quite generally our conditions accurately predict the conditions under which fragmentation occurs in numerical simulations.

In the next section we will apply our analytical conditions to various examples of inflation models and compare them with the known numerical results in order to demonstrate the effectiveness of our analytic fragmentation conditions.

%%%%%%%%%%%%%%%%%%%%%%%%
\section{Applications}
\label{sec:apps}
%%%%%%%%%%%%%%%%%%%%%%%%

%%%%%%%%%%%%%%%%%%%%%%%%
\subsection{$\alpha$-attractor $T$-model}
\label{subsec:alphaattractorT}
%%%%%%%%%%%%%%%%%%%%%%%%

The $\alpha$-attractor $T$-model \cite{Kallosh:2013yoa,Galante:2014ifa,Kallosh:2015lwa} with $n=1$ has the following potential in the Einstein frame:
\begin{align}
V = \lambda \tanh^2\left(\frac{\Phi}{\sqrt{6\alpha}M_{\rm P}}\right)\,.
\end{align}
Around the minimum, the potential can be expanded as
\begin{align}
V \approx \frac{\lambda}{6\alpha}\left(\frac{\Phi}{M_{\rm P}}\right)^2
-\frac{\lambda}{54\alpha^2}\left(\frac{\Phi}{M_{\rm P}}\right)^4\,.
\end{align}
We thus find, by comparing with Eq. \eqref{eqn:PotSymm},
\begin{align}
m = \sqrt{\frac{\lambda}{3\alpha M_{\rm P}^2}}
\,,\quad
A = \frac{\lambda}{54\alpha^2 M_{\rm P}^4}\,.
\end{align}
Therefore, by substituting $m$ and $A$ into Eq. \eqref{eqn:MainResultSymm}, we see that the fragmentation condition is satisfied if
\begin{align}
\alpha \lesssim 5.6 \times 10^{-5}\left(\frac{r_S}{0.1}\right)\,.
\end{align}
Thus, in the $\alpha$-attractor $T$-model, fragmentation of the inflaton condensate will occur for $\alpha \lesssim 5\times10^{-5}$. Our analytical result is in agreement with the numerical analysis of Ref. \cite{Lozanov:2016hid}, which demonstrates fragmentation for an explicit example with $\alpha \approx 10^{-5}$, but finds no fragmentation for larger values 
of $\alpha$. A recent analytical and numerical study of tachyonic preheating in plateau inflation models has also obtained a similar bound \cite{Tomberg:2021bll,Koivunen:2022mem}.

%%%%%%%%%%%%%%%%%%%%%%%%
\subsection{$\alpha$-attractor $E$-model}
\label{subsec:alphaattractorE}
%%%%%%%%%%%%%%%%%%%%%%%%

The Einstein-frame potential of the $\alpha$-attractor $E$-model \cite{Kallosh:2013yoa,Galante:2014ifa,Kallosh:2015lwa} with $n=1$ is given by
\begin{align}\label{eqn:alphaEpot}
V = \lambda \left[
1 - \exp\left(
-\sqrt{\frac{2}{3\alpha}}\frac{\Phi}{M_{\rm P}}
\right)
\right]^2\,.
\end{align}
The potential can be expanded around the minimum as
\begin{align}
V \approx \frac{2\lambda\Phi^2}{3\alpha M_{\rm P}^2}
-\frac{2}{3}\sqrt{\frac{2}{3}}\frac{\lambda}{\alpha^{3/2}}\frac{\Phi^3}{M_{\rm P}^3}\,.
\end{align}
Comparing with Eq. \eqref{eqn:PotAsymm}, we find
\begin{align}
m = \sqrt{\frac{4\lambda}{3\alpha M_{\rm P}^2}}
\,,\quad
A = \frac{2}{3}\sqrt{\frac{2}{3}}\frac{\lambda}{\alpha^{3/2}M_{\rm P}^3}\,.
\end{align}
Therefore, by substituting $m$ and $A$ into Eq. \eqref{eqn:MainResultAsymm}, we see that the fragmentation condition is satisfied if
\begin{align}\label{eqn:alphaEcond}
\alpha \lesssim 5\times 10^{-3}\left(\frac{r_A}{0.1}\right)^2\,.
\end{align}
Thus, in the $\alpha$-attractor $E$-model, fragmentation of the inflaton condensate will occur for $\alpha \lesssim 5\times 10^{-3}$. This analytical result agrees with a numerical analysis performed in Ref. \cite{Hasegawa:2017iay}, where it is concluded that fragmentation will occur for $\alpha \lesssim 10^{-3}$.

%%%%%%%%%%%%%%%%%%%%%%%%
\subsection{Starobinsky $R^2$ model}
\label{subsec:R2model}
%%%%%%%%%%%%%%%%%%%%%%%%
Starobinsky's $R^2$ inflation model \cite{Starobinsky:1980te} has the following potential in the Einstein frame,
\begin{align}
V = \lambda \left[ 1 - \exp\left( -\sqrt{\frac{2}{3}} \frac{\Phi}{M_{\rm P}} \right) \right]^2\,.
\end{align}
We note that this potential can be obtained from the $\alpha$-attractor $E$-model potential \eqref{eqn:alphaEpot} by setting $\alpha = 1$. From Eq. \eqref{eqn:alphaEcond}, we see that $\alpha = 1$ does not satisfy the fragmentation condition. Therefore, we conclude that the fragmentation does not occur in the $R^2$ model. This is consistent with the results of Ref. \cite{Takeda:2014qma}.

%%%%%%%%%%%%%%%%%%%%%%%%
\subsection{Palatini $R^2$ model with a quadratic potential}
\label{subsec:phi2R2model}
%%%%%%%%%%%%%%%%%%%%%%%%

Let us consider the quadratic inflation model in the Palatini formulation with an $R^2$ term added \cite{Enckell:2018hmo,Karam:2021sno}. The action is given in the Jordan frame by 
\begin{align}
S &= \int d^4x \, \sqrt{-g_{\rm J}} \, \bigg[
\frac{M_{\rm P}^2}{2}g_{\rm J}^{\mu\nu}R_{{\rm J} \mu\nu}
+ \frac{\alpha^2}{4}(g_{\rm J}^{\mu\nu}R_{{\rm J} \mu\nu})^2
\nonumber\\
&\qquad\qquad\qquad\qquad
-\frac{1}{2}g^{\mu\nu}_{\rm J}\partial_\mu\varphi\partial_\nu\varphi
-\frac{1}{2}m^2\varphi^2
\bigg]\,,
\end{align}
where we put the subscript J to denote the Jordan frame.
The action can be brought into the Einstein frame via Weyl rescaling\footnote{
As our aim is to apply our general analytical conditions to a specific model, we do not discuss the detailed computational steps. Readers may refer to Refs. \cite{Enckell:2018hmo,Karam:2021sno}.
},
\begin{align}
S = \int d^4 x \, \sqrt{-g_{\rm E}} \, \left[
\frac{M_{\rm P}^2}{2}g_{\rm E}^{\mu\nu}R_{{\rm E} \mu\nu}
-\frac{1}{2}g^{\mu\nu}_{\rm E}\partial_\mu\Phi\partial_\nu\Phi
-V
\right]\,,
\end{align}
where the subscript E stands for the Einstein frame, $\Phi$ is the canonically normalised field, and $V$ is the Einstein-frame potential.
Near $\varphi = 0$, the canonically normalised field can be approximated as $\Phi \approx \varphi$, and the Einstein-frame potential is given by \cite{Lloyd-Stubbs:2020pvx}
\begin{align}
V \approx \frac{1}{2}m^2\Phi^2 - \alpha\left(
\frac{m}{M_{\rm P}}
\right)^4\Phi^4\,.
\end{align}
Thus, it belongs to the symmetric case \eqref{eqn:PotSymm} with
\begin{align}
A = \alpha\left(\frac{m}{M_{\rm P}}\right)^4\,.
\end{align}
Substituting the expression of $A$ into Eq. \eqref{eqn:MainResultSymm}, we see that the fragmentation condition is satisfied if
\begin{align}
\alpha\left(\frac{m}{M_{\rm P}}\right)^2 > 1000\left(\frac{0.1}{r_S}\right)\,.
\end{align}
To match the magnitude of the primordial curvature power spectrum, $2 \times 10^{-9}$, the inflaton mass needs to be $m \simeq 1.4 \times 10^{13}$ GeV. The prediction for the fragmentation condition is then \cite{Lloyd-Stubbs:2020pvx}
\begin{align}
\alpha \gtrsim 2.94 \times 10^{13}
\left(\frac{0.1}{r_S}\right)
\,.
\end{align}
The same lower bound of $10^{13}$ was later found in Ref. \cite{Karam:2021sno} by performing a numerical analysis of tachyonic preheating in this model, and more recently by an analytical and numerical study in Ref. \cite{Tomberg:2021bll}.

%%%%%%%%%%%%%%%%%%%%%%%%
\subsection{Higgs Inflation}
\label{subsec:HImodel}
%%%%%%%%%%%%%%%%%%%%%%%%
In the previous sections we considered models for which a numerical analysis of fragmentation already exists, and we demonstrated that our simple analytical conditions can reproduce the conditions for fragmentation that were previously obtained numerically. In this section, we consider examples for which no numerical analysis at present exists, namely Higgs Inflation with a symmetry-breaking potential in both the metric and Palatini formalisms, where the inflaton is oscillating around the symmetry-breaking minimum of its potential. We derive the analytical conditions for fragmentation and then confirm numerically that these conditions are correct.

Higgs Inflation is described the following action in the Jordan frame:
\begin{align}
S &= \int d^4x \, \sqrt{-g_{\rm J}} \, \bigg[
\frac{M_{\rm P}^2}{2}F(\varphi)g_{\rm J}^{\mu\nu}R_{{\rm J}\mu\nu}
\nonumber\\
&\qquad\qquad\qquad\qquad
-\frac{1}{2}g_{\rm J}^{\mu\nu}\partial_\mu\varphi\partial_\nu\varphi
-V_{\rm J}(\varphi)
\bigg]\,,
\end{align}
where the subscript J denotes the Jordan frame, $F(\varphi) = (M^2 + \xi\varphi^2)/M_{\rm P}^2$ is the non-minimal coupling term with $M$ being a mass parameter, which reproduces the Planck mass at today, i.e., $M^2 + \xi v^2 \equiv M_{\rm P}^2$, $v$ is the vacuum expectation value of the $\varphi$ field, and $V_{\rm J}(\varphi) = \lambda(\varphi^2 - v^2)^2/4$ is the Higgs potential in the Jordan frame.
In the Einstein frame, the action is given by
\begin{align}
S = \int d^4x \,\sqrt{-g_{\rm E}}\left[
\frac{M_{\rm P}^2}{2}g_{\rm E}^{\mu\nu}R_{{\rm E}\mu\nu}
- \frac{1}{2}g_{\rm E}^{\mu\nu}\partial_\mu\Phi\partial_\nu\Phi
-V_{\rm E}(\varphi)
\right]\,,
\end{align}
where the subscript E denotes the Einstein frame, $\Phi$ is the canonically normalised field, and $V_{\rm E}$ is the scalar potential in the Einstein frame. The Einstein-frame field and potential are related to the Jordan-frame field and potential as
\begin{align}
\frac{d\Phi}{d\varphi} &=
\sqrt{\frac{1}{F} + \sigma \frac{3M_{\rm P}^2}{2F^2}\left(\frac{dF}{d\varphi}\right)^2}
\,,\label{eqn:dPhidvarphi}\\
V_{\rm E} &= \frac{V_{\rm J}}{F^2}\,,
\end{align}
where $\sigma$ parametrises the metric ($\sigma = 1$) and the Palatini ($\sigma = 0$) formulations. We shall omit the subscript E in the following.
For inflationary physics of the Higgs Inflation in the metric and the Palatini formulations, see, e.g., Refs. \cite{Bauer:2008zj,Takahashi:2018brt,Rubio:2019ypq}.

Assuming that $|\varphi - v| \ll v$, we can expand $V(\varphi)$ as
\begin{align}
V(\varphi) &=
\frac{\lambda}{4F^2}(\varphi+v)^2(\varphi-v)^2
\nonumber\\
&=
\frac{\lambda}{4F^2}[(\varphi-v)+2v]^2(\varphi-v)^2
\nonumber\\
&\approx
\frac{\lambda v^2}{F^2}(\varphi - v)^2\left(
1 + \frac{\varphi - v}{v}
\right)\,.
\end{align}
We can also expand $F$ as
\begin{align}
M_{\rm P}^2F 
&=M_{\rm P}^2 + \xi(\varphi+v)(\varphi-v)
\nonumber\\
&=M_{\rm P}^2 + \xi[2v+(\varphi-v)](\varphi-v)\,,
\end{align}
and thus we find
\begin{align}
\frac{1}{F^2} \approx
1 - \frac{4\xi v}{M_{\rm P}^2}(\varphi-v)
-\frac{2\xi}{M_{\rm P}^2}\left(
1 - \frac{6\xi v^2}{M_{\rm P}^2}
\right)(\varphi-v)^2\,.
\end{align}
Thus, to leading order in $(\varphi-v)$, the potential $V(\varphi)$ can be expressed as
\begin{align}
V(\varphi) \approx \lambda v^2 (\varphi-v)^2 \left[
1 - \left(
\frac{4\xi v^2}{M_{\rm P}^2} - 1
\right)\frac{\varphi-v}{v}
\right]\,.
\end{align}

For $|\varphi - v| \ll v$, the relation between $\varphi$ and $\Phi$ becomes
\begin{align}
\Phi \approx \sqrt{1+\frac{6\sigma \xi^2 v^2}{M_{\rm P}^2}}(\varphi-v)\,,
\end{align}
where we define $\Phi = 0$ at $\varphi = v$. This follows since, to leading order in $(\varphi-v)$, we can set $\varphi = v$ on the right-hand side of Eq. \eqref{eqn:dPhidvarphi}. Thus, in terms of $\Phi$, the Einstein-frame potential for $|\varphi - v|\ll v$ becomes
\begin{align}
V(\Phi) = \frac{\lambda v^2}{1+6\sigma\xi^2v^2/M_{\rm P}^2}\Phi^2 \mp \frac{\lambda v|4\xi v^2/M_{\rm P}^2 - 1|}{(1+6\sigma\xi^2v^2/M_{\rm P}^2)^{3/2}}\Phi^3\,.
\end{align}
This is of the form of Eq. \eqref{eqn:PotAsymm} with
\begin{align}
m^2 = \frac{2\lambda v^2}{1+6\sigma\xi^2v^2/M_{\rm P}^2}
\,,\quad
A = \frac{\lambda v|4\xi v^2/M_{\rm P}^2 - 1|}{(1+6\sigma\xi^2v^2/M_{\rm P}^2)^{3/2}}\,.
\end{align}
Applying the fragmentation condition for an asymmetric potential, \eqref{eqn:MainResultAsymm}, we obtain the general condition for fragmentation in Higgs Inflation with a symmetric-breaking potential as follows:
\begin{align}
\frac{v}{M_{\rm P}} \lesssim
\frac{9\pi}{400}\sqrt{\frac{3}{2}}\left(\frac{r_A}{0.1}\right)
\frac{|4\xi v^2/M_{\rm P}^2 - 1|}{(1+6\sigma\xi^2v^2/M_{\rm P}^2)^{1/2}}\,.
\end{align}

\underline{(i) Case $M^2 \ll \xi v^2$:} In this case, $M_{\rm P}^2 \approx \xi v^2$, corresponding to non-minimally coupled inflation in the induced gravity limit. In this limit, the condition for fragmentation becomes
\begin{align}
\frac{(1+6\sigma\xi)^{1/2}}{\sqrt{\xi}} \lesssim
0.26\left(\frac{r_A}{0.1}\right)\,.
\end{align}
For the metric case, $\sigma = 1$, and, assuming that $\xi \gg 1$, the condition becomes
\begin{align}
\sqrt{6} \lesssim 0.26\left(\frac{r_A}{0.1}\right)\,,
\end{align}
which is not satisfied. Therefore, fragmentation does not occur for metric Higgs Inflation in the induced gravity limit. For the Palatini case, $\sigma = 0$, and the condition becomes
\begin{align}
\xi \gtrsim 14.83
\left(\frac{0.1}{r_A}\right)^2\,.
\end{align}
This is easily satisfied since we expect that $\xi \gg 1$. Therefore, we expect that the inflaton condensate will fragment in the case of Palatini Higgs Inflation in the induced gravity limit, but it will not fragment in the case of metric Higgs Inflation in this limit.

\underline{(ii) Case $M^2 \gg \xi v^2$:} In this case, $M_{\rm P}^2 \approx M^2$, corresponding to conventional Higgs Inflation. In addition, since $|\varphi - v|\ll v$, we also have $\xi\varphi^2 \ll M_{\rm P}^2$. In this limit, the condition for fragmentation becomes
\begin{align}
\frac{v}{M_{\rm P}}\lesssim
0.087\left(\frac{0.1}{r_A}\right)\left(
1+\frac{6\sigma\xi^2v^2}{M_{\rm P}^2}\right)^{-1/2}\,.
\end{align}
For the metric case, $\sigma = 1$, and the condition becomes
\begin{align}
\frac{v}{M_{\rm P}}\left(
1+\frac{6\xi^2v^2}{M_{\rm P}^2}
\right)^{1/2} \lesssim 0.087\left(\frac{0.1}{r_A}\right)\,.
\end{align}
There are two cases, corresponding to $6\xi^2v^2/M_{\rm P}^2 \ll 1$ and $6\xi^2 v^2/M_{\rm P}^2 \gg 1$. In the case where $6\xi^2v^2/M_{\rm P}^2 \ll 1$, the condition for the fragmentation becomes
\begin{align}
\frac{v}{M_{\rm P}} \lesssim 0.087\left(
\frac{0.1}{r_A}\right)\,.
\end{align}
In this case, the non-minimal coupling plays no role and we simply have the condition for fragmentation for a minimally coupled scalar with a broken-symmetry potential. This is satisfied as long as $v$ is not very close to the Planck scale.

In the case where $6\xi^2 v^2/M_{\rm P}^2 \gg 1$, the condition for fragmentation becomes
\begin{align}
\frac{\xi v^2}{M_{\rm P}^2} \lesssim
0.035\left(\frac{0.1}{r_A}\right)\,.
\end{align}
In this case, the non-minimal coupling plays a role, even though $\xi\varphi^2 \ll M_{\rm P}^2$. This is because once $\varphi>M_{\rm P}/(\sqrt{6}\xi)$, the inflaton kinetic term in the Einstein frame in the metric case is modified. The allowed range of $\xi v^2/M_{\rm P}^2$ is $1 \gg \xi v^2/M_{\rm P}^2 \gg 1/(6\xi)$. Therefore, the fragmentation condition will be satisfied for most of the allowed range of $\xi v^2/M_{\rm P}^2$ except for when $\xi v^2$ approaches $M_{\rm P}^2$.

In the case of Palatini Higgs Inflation, $\sigma = 0$, and the fragmentation condition becomes
\begin{align}
\frac{v}{M_{\rm P}} \lesssim 0.087\left(\frac{0.1}{r_A}\right)\,.
\end{align}
This is simply the fragmentation condition for a minimally coupled scalar with a broken-symmetry potential. This is expected as in the limit $\xi v^2 \ll M_{\rm P}^2$ (and so $\xi\varphi^2 \ll M_{\rm P}^2$), the non-minimal coupling plays no role in Palatini inflation.

\begin{figure*}[ht!]
\centering
\begin{minipage}[b]{.4\textwidth}
\includegraphics[width=1.0\textwidth]{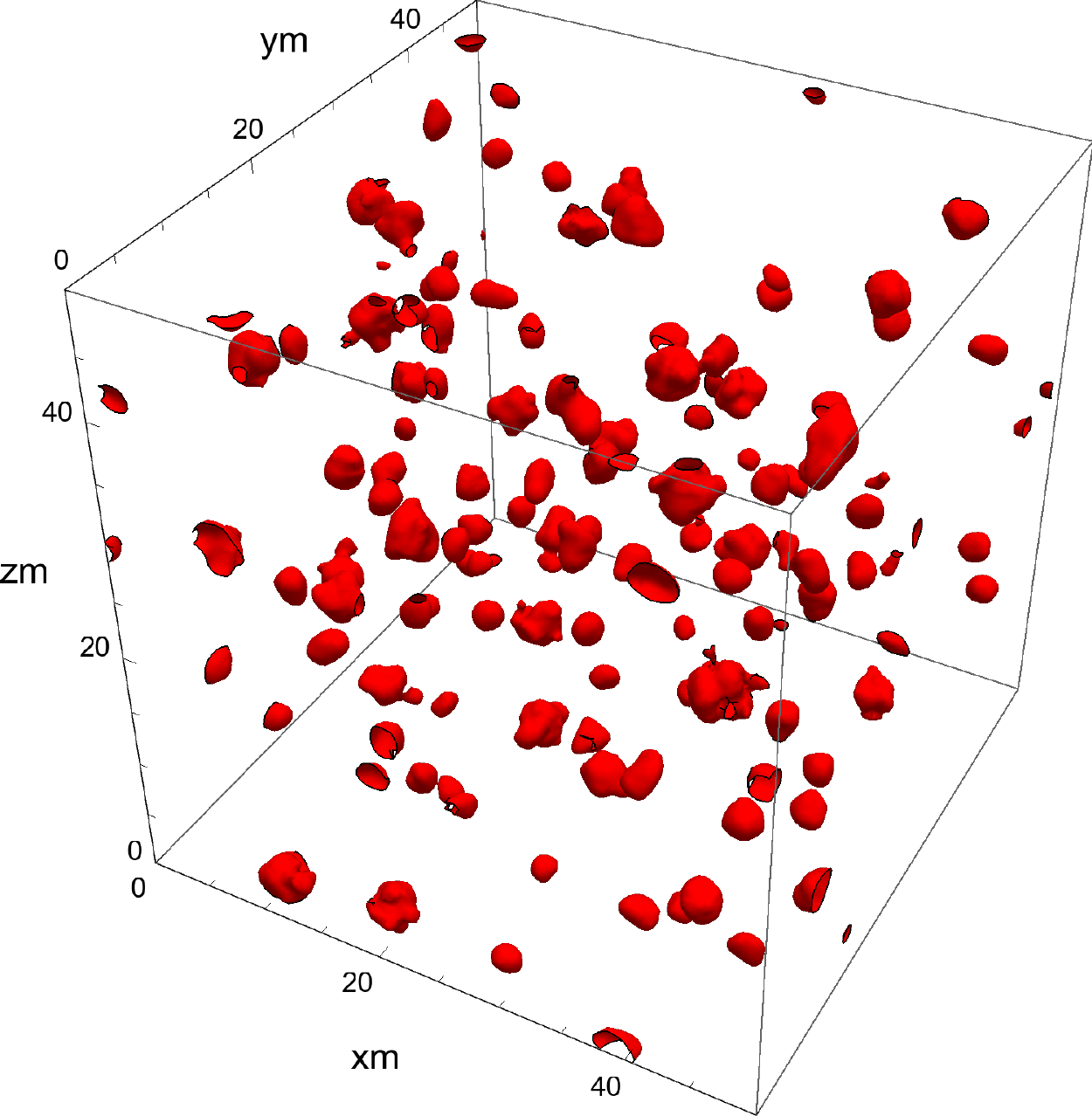}
\caption{Lattice simulation for the Palatini Higgs Inflation model. We present the quantity $\rho/\langle\rho\rangle = 10$ at $t=500m^{-1}$. The choice of parameters are as follows: $N_{{\rm grid}} = 128^{3}$, $L=50m^{-1}$, $\xi = 1.0\times 10^{3}$, and $\sqrt{\xi} v/M = 1.0 \times 10^{3}$. We used the field value at the end of inflation, $\Phi_{{\rm end}} = (M_{{\rm P}}/2\sqrt{\xi})\sinh^{-1}(4\sqrt{2\xi})$, as the initial condition.}\label{fig:LE1D-largexiv}
\end{minipage}\qquad\qquad\qquad
\begin{minipage}[b]{.4\textwidth}
\includegraphics[width=1.0\textwidth]{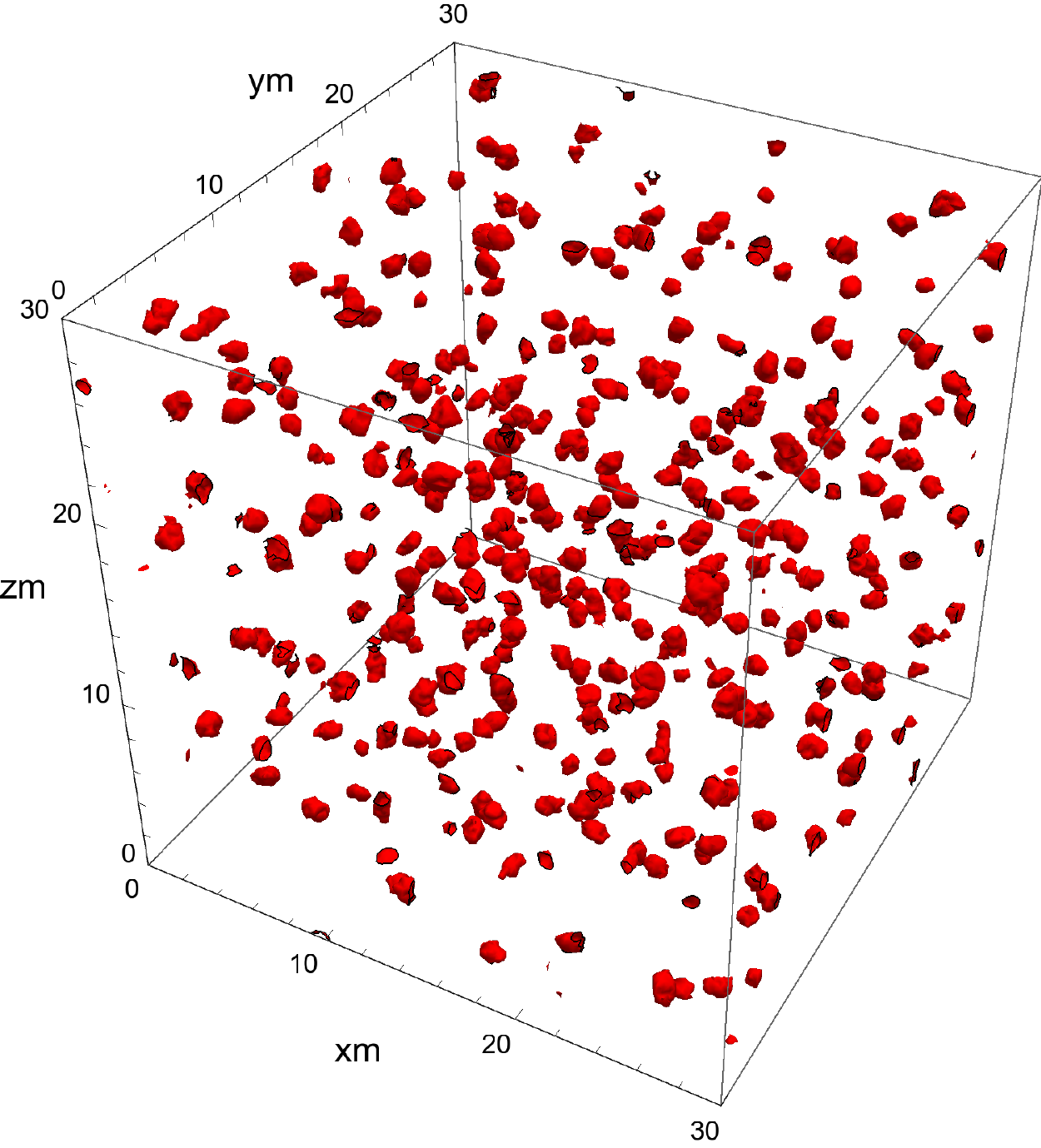}
\caption{Lattice simulation for the Higgs Inflation model. We present the quantity $\rho/\langle\rho\rangle = 10$ at $t=40m^{-1}$. The choice of parameters are as follows: $N_{{\rm grid}} = 128^{3}$, $L=30m^{-1}$, $\xi = 1.0\times 10^{4}$ and $\sqrt{\xi}v/M = 0.1$. We used the field value at the end of inflation, $\Phi_{{\rm end}} = (M_{{\rm P}}/2\sqrt{\xi})\sinh^{-1}(4\sqrt{2\xi})$, as the initial condition.}\label{fig:LE1D-smallxiv}
\end{minipage}
\end{figure*}

In summary, we find significantly new results for the case of metric and Palatini Higgs Inflation with a broken-symmetry potential in the induced gravity limit, and for metric Higgs Inflation with a broken-symmetry potential in the limit where $v>M_{\rm P}/(\sqrt{6}\xi)$, with all other cases reducing to a conventional minimally coupled scalar with a symmetry-breaking potential. In particular, we find that fragmentation will occur for Palatini Higgs Inflation in the induced gravity limit, but will not occur for metric Higgs Inflation in this limit.

To our knowledge, there is no numerical study of this particular setup. In order to verify our analytical study, we perform numerical simulations in 1+3 dimensions by using the public lattice simulation code {\tt LATTICEEASY}~\cite{Felder:2000hq}.
We present results of numerical simulations for the following two cases:
\begin{itemize}
\item Palatini Higgs inflation in the $\xi v^{2} \gg M^{2}$ limit and
\item Metric and Palatini Higgs inflation with $\xi v^{2} \ll M^{2}$, together with the condition $\xi^{2}v^{2} \ll M^{2}$.
\end{itemize}
In both cases, our analytical results predict that fragmentation will occur. Note that, in the second case, the fragmentation conditions are same in both formalisms as in this limit both models reduce to a minimally coupled scalar with a symmetry-breaking potential, and there is no difference between the formalisms.

For the numerical simulation, we set $N_{{\rm grid}} = 128^{3}$, $L=50m^{-1}$ ($30m^{-1}$), and $\delta t = 0.1m^{-1}$ for the first (second) case, where $m \equiv \sqrt{2\lambda}v$.
For both cases, we used the field value at the end of inflation, $\Phi_{{\rm end}} = (M_{{\rm P}}/2\sqrt{\xi})\sinh^{-1}(4\sqrt{2\xi})$ \cite{Takahashi:2018brt}, as the initial condition.
The result of the numerical simulation for the case of $\xi v^{2}/M^{2} \gg 1$ is shown in Fig.~\ref{fig:LE1D-largexiv}. We present the energy density $\rho/ \langle\rho\rangle = 10$, where $\langle\rho\rangle$ is the averaged energy density over the lattice.
In Fig.~\ref{fig:LE1D-smallxiv}, we plotted the energy density $\rho/\langle\rho\rangle = 10$ for the case of $\xi v^{2}/M^{2} \ll 1$.
We see that the inflaton condensate fragments after inflation ends in both cases.
Our analytical results are thus in agreement with the numerical analysis.

%%%%%%%%%%%%%%%%%%%%%%%%
\section{Conclusion}
\label{sec:conc}
%%%%%%%%%%%%%%%%%%%%%%%%
In this paper, we have derived general analytical conditions under which the inflaton condensate will fragment for the case of both symmetric and asymmetric potentials.
The robustness of our results was demonstrated by applying our analytical fragmentation conditions to a range of models for which the result is known numerically, including the $\alpha$-attractor $T$ and $E$ models, Starobinsky's $R^2$ model, and the Palatini $R^2$ model with a quadratic potential. In all cases, we find that the analytically predicted condition on the model parameters for fragmentation to occur are in complete agreement with the results of the numerical analyses.

In addition, we have applied our results to Higgs Inflation with a broken-symmetry potential in both the metric and Palatini formulations and derived a general condition for fragmentation to occur in these models. We then carried out numerical simulations of these models using {\tt LATTICEEASY} and found complete agreement between our analytical predictions for fragmentation to occur and the results of the numerical simulations.

The conditions we have derived provide a quick and simple method to check whether any model which can support non-topological soliton solutions (which requires that the scalars have a mass at the potential minimum and therefore that the potential is approximately quadratic at the minimum) will undergo
fragmentation and to determine the range of model parameters for which this is possible.

The physics of the inflaton condensate fragmentation has many interesting phenomenological consequences, including the formation of primordial black holes and gravitational wave signals. The evolution dynamics generally requires extensive numerical lattice simulations. We believe that our findings may serve as a starting point for such numerical analyses.

%%%%%%%%%%%%%%%%%%%%%%%%
\section*{Acknowledgements}
%%%%%%%%%%%%%%%%%%%%%%%%
J.K. would like to thank Kaloian D. Lozanov for useful discussion on the use of {\tt LATTICEEASY}.

%%%%%%%%%%%%%%%%%%%%%%%%

%%%%%%%%%%%%%%%%%%%%%%%%
\end{document}